\documentclass[prd,final,superscriptaddress,preprintnumbers,eqsecnum,showpacs,nofootinbib,nobibnotes,noeprint,twocolumn]{revtex4-2}

\usepackage{amssymb}
\usepackage{lipsum}
\usepackage{color}
\usepackage{graphicx}
\usepackage{dcolumn}
\usepackage{bm}
\usepackage[utf8]{inputenc}
\usepackage{gensymb}
\usepackage{latexsym}
\usepackage{amsmath}
\usepackage{amsfonts}
\usepackage{nicefrac}
\usepackage{float}
\usepackage{booktabs}
\usepackage{siunitx}
\usepackage{slashed}
\usepackage{hhline}
\usepackage[table]{xcolor}
\usepackage[mathscr,scaled=1.15]{urwchancal}

\usepackage{url}
\usepackage{xspace}
\usepackage{siunitx}
\usepackage{xfrac}
\usepackage{hyperref}
\usepackage[nameinlink]{cleveref}
\usepackage{appendix}

\usepackage{xifthen}
\usepackage{xcolor}
\hypersetup{colorlinks,	linkcolor={blue},	citecolor={blue},	urlcolor={blue}}

\DeclareFontFamily{OT1}{pzc}{}
\DeclareFontShape{OT1}{pzc}{m}{it}%
{<-> s * [1.15] pzcmi7t}{}
\DeclareMathAlphabet{\mathpzc}{OT1}{pzc}{m}{it}

\newcommand{\be}{\begin{equation}}
	\newcommand{\bea}{\begin{eqnarray}}
		\newcommand{\ee}{\end{equation}}
	\newcommand{\eea}{\end{eqnarray}}

\def\1eq#1{Eq.~(\ref{#1})}

\def\2eqs#1#2{Eqs.~(\ref{#1}) and (\ref{#2})}

\graphicspath{{./figures/}{./}}

\begin{document}

\title{Hadronic vacuum polarization contribution to $a_\mu$ from functional methods with strong and electromagnetic isospin breaking}

\author{Angel S. Miramontes}
 \email{angel.s.miramontes@uv.es}
\affiliation{Department of Theoretical Physics and IFIC, University of Valencia and CSIC, E-46100, Valencia, Spain}

\author{Adnan Bashir}
\affiliation{Departamento de Ciencias Integradas, Universidad de Huelva, E-21071 Huelva, Spain.}
\affiliation{Facultad de Ingenier\'ia, Universidad Aut\'onoma de Quer\'etaro, Quer\'etaro, Quer\'etaro 76010, M\'exico}

\author{Christian S. Fischer}
\affiliation{Institut für Theoretische Physik, Justus-Liebig-Universität Gießen, 35392 Gießen, Germany}
\affiliation{Helmholtz Forschungsakademie Hessen für FAIR (HFHF),
GSI Helmholtzzentrum für Schwerionenforschung, Campus Gießen, Gießen, 35392 Germany}

\author{Pablo Roig}
\affiliation{Departamento de Física, Centro de Investigación y de Estudios
	Avanzados del Instituto Politécnico Nacional, Apartado Postal 14-740,
	07360, Mexico City, Mexico.}

\begin{abstract}
We present a continuum-QCD determination of the leading-order hadronic vacuum-polarization contribution to the anomalous magnetic moment of the muon within the Dyson-Schwinger and Bethe-Salpeter equation framework. The calculation incorporates pion back-reaction, a fully dressed quark-photon vertex with a dynamically generated $\rho$-resonance structure in the timelike region, and both strong and electromagnetic isospin breaking treated self-consistently at the quark level. We obtain
$a_\mu^{\mathrm{HVP,LO}}(u+d+s+c)|_{\mathrm{ISB}} = 709.7 \times 10^{-10}$, 
in good agreement with recent lattice-QCD determinations, and find an isospin-breaking shift of $\Delta a_\mu^{\mathrm{HVP,LO}} = 4.5 \times 10^{-10}$ ($0.6\%$), demonstrating that isospin-breaking effects, while quantitatively modest, are not negligible. Including the bottom-quark contribution and an indicative estimate of the systematic uncertainties, we obtain our final result, $a_\mu^{\mathrm{HVP,LO}}(u+d+s+c+b)|_{\mathrm{ISB}} = (710.0 \pm 14.5) \times 10^{-10}$.

\end{abstract}

\maketitle

\section{Introduction}
 The anomalous magnetic moment of the muon, $a_\mu$, is a uniquely sensitive probe of the Standard Model (SM). 
The uncertainty in its SM prediction,~\cite{Aliberti:2025beg}\footnote{This result is based on Refs.~\cite{Aoyama:2012wk,Volkov:2019phy,Volkov:2024yzc,Aoyama:2024aly,Parker:2018vye,Morel:2020dww,Fan:2022eto,Czarnecki:2002nt,Gnendiger:2013pva,Ludtke:2024ase,Hoferichter:2025yih,RBC:2018dos,Giusti:2019xct,Borsanyi:2020mff,Lehner:2020crt,Wang:2022lkq,Aubin:2022hgm,Ce:2022kxy,ExtendedTwistedMass:2022jpw,RBC:2023pvn,Kuberski:2024bcj,Boccaletti:2024guq,Spiegel:2024dec,RBC:2024fic,Djukanovic:2024cmq,ExtendedTwistedMass:2024nyi,MILC:2024ryz,FermilabLatticeHPQCD:2024ppc,Keshavarzi:2019abf,DiLuzio:2024sps,Kurz:2014wya,Colangelo:2015ama,Masjuan:2017tvw,Colangelo:2017fiz,Hoferichter:2018kwz,Eichmann:2019tjk,Bijnens:2019ghy,Leutgeb:2019gbz,Cappiello:2019hwh,Masjuan:2020jsf,Bijnens:2020xnl,Bijnens:2021jqo,Danilkin:2021icn,Stamen:2022uqh,Leutgeb:2022lqw,Hoferichter:2023tgp,Hoferichter:2024fsj,Estrada:2024cfy,Deineka:2024mzt,Eichmann:2024glq,Bijnens:2024jgh,Hoferichter:2024bae,Holz:2024diw,Cappiello:2025fyf,Colangelo:2014qya,Blum:2019ugy,Chao:2021tvp,Chao:2022xzg,Blum:2023vlm,Fodor:2024jyn}.}, is completely dominated by the error budget of the  
hadronic vacuum polarization (HVP). 
The leading-order HVP contribution, denoted as
$a_\mu^{\mathrm{HVP,LO}}$, can be determined using several complementary approaches. The lattice-QCD calculations have made substantial progress in recent years, reducing the uncertainty in the determination of this observable to $0.87\%$~\cite{Aliberti:2025beg}. On the other hand, there is a hybrid
strategy which combines lattice-QCD information with the measured
$e^+e^-\to\mathrm{hadrons}$ cross section through dispersion relations. It exploits the respective strengths of lattice calculations and experimental data in different kinematic regions. It has achieved a precision of $0.48\%$~\cite{Boccaletti:2024guq}. Moreover, traditional dispersive
evaluations based entirely on $e^+e^-\to\mathrm{hadrons}$ data remain an attractive alternative, but are presently limited by tensions among some of the available data sets~\cite{Aliberti:2025beg}. Hadronic spectral functions
extracted from $\tau\to\pi\pi\nu_\tau$ decays offer a further complementary
source of information, provided that the required strong and electromagnetic isospin-breaking corrections are brought under adequate control~\cite{Miranda:2020wdg,Masjuan:2023qsp,Davier:2023fpl,Castro:2024prg,
Colangelo:2025iad,Colangelo:2025ivq,Cirigliano:2026ios,Allen:2026iad,Bruno:2026yba}. The need to
resolve these issues has become more pressing following the Fermilab
measurement: the present experimental world average for $a_\mu$ has an
uncertainty approximately four times smaller than that of the
SM prediction~\cite{Muong-2:2025xyk,Muong-2:2023cdq,Muong-2:2024hpx,Muong-2:2021ojo,
Muong-2:2021vma,Muong-2:2021ovs,Muong-2:2021xzz,Muong-2:2006rrc,
Aliberti:2025beg}. Therefore, further progress in the theoretical determination of
$a_\mu^{\mathrm{HVP,LO}}$ is essential to fully exploit this
experimental precision in stringent tests of the Standard Model.

A major achievement of the theory initiative on $a_\mu$~\cite{Aoyama:2020ynm} has been to bring together practitioners employing complementary nonperturbative approaches, including dispersive methods, lattice QCD, effective field theory, holographic QCD, and functional techniques. These methods play distinct but mutually reinforcing roles. Lattice-QCD and dispersive approaches are, in principle, capable of delivering high-precision determinations with systematically controlled uncertainty analyses, whereas functional methods are particularly valuable in two respects: first, they provide independent cross-checks of contributions obtained using other frameworks; second, they yield quantitative exploratory estimates for contributions that have not yet been determined reliably by alternative methods, with uncertainties that can be assessed, albeit not always through the same systematically improvable procedures available in lattice-QCD or dispersive approaches.

This complementarity was already evident in the first theory white paper~\cite{Aoyama:2020ynm}. Functional calculations reproduced the central values of the pion-exchange and pion-loop contributions to hadronic light-by-light (HLbL) scattering; they also supplied important cross-checks of Canterbury-extrapolant determinations of the $\eta$- and $\eta'$-exchange contributions, and provided the only available estimate of the kaon-loop contribution \cite{Eichmann:2019bqf,Aoyama:2020ynm,Miramontes:2021exi}. This latter result was subsequently corroborated by dispersive analyses \cite{Stamen:2022uqh,Aliberti:2025beg}. To date, the principal focus of the functional approaches has been to compute the HLbL contributions, Refs.~\cite{Goecke:2010if,Raya:2019dnh,Eichmann:2019tjk,Eichmann:2019bqf,Eichmann:2024glq,Miramontes:2021exi,Miramontes:2024fgo,Xing:2025iab}, whereas calculations of the hadronic vacuum-polarization contribution have remained exploratory~\cite{Goecke:2011pe}. The primary aim of the present work is to advance the functional determination of HVP to a robust quantitative level.

Since the present calculation is formulated within the continuum functional approach to QCD, we begin with a few general remarks on the coupled formalism based on the Dyson-Schwinger and Bethe-Salpeter equations (DSE/BSE). This framework provides a unified description of QCD Green's functions as well as hadronic observables, yielding direct access to the dressed quark propagator and quark-photon vertex, the fundamental building blocks of the hadronic vacuum polarization (HVP) tensor.
The DSE/BSE approach has been extensively employed in studies of hadron spectroscopy and structure, enabling systematic investigations of mesons, baryons, and their electromagnetic properties \cite{Maris:1997hd,Maris:1999bh,Sanchis-Alepuz:2014sca,Sanchis-Alepuz:2014wea,Sanchis-Alepuz:2013iia,Sanchis-Alepuz:2015qra,Raya:2015gva,Raya:2016yuj,Eichmann:2016hgl,Eichmann:2016yit, Bashir:2012fs, Sanchis-Alepuz:2017mir,Williams:2018adr,Ding:2018xwy,Miramontes:2021xgn,Miramontes:2022uyi,Chen:2023zhh,Liu:2023reo,Miramontes:2025ofw,Miramontes:2025vzb,Ferreira:2025wpu, Hagel:2025ngi,Ferreira:2026gbe}. In recent years, the same tools have also been extended to address more complex bound states such as four-quark states~\cite{Heupel:2012ua,Wallbott:2019dng,Santowsky:2020pwd,Hoffer:2024fgm,Hoffer:2024alv} and glueballs~\cite{Sanchis-Alepuz:2015hma, Huber:2020ngt,Huber:2021yfy,Huber:2025kwy}, illustrating the breadth of the framework and the dynamical information encoded in its Green's functions. 
In the context of HVP they offer a complementary and independent continuum perspective including the possibility to assess the impact of strong and electromagnetic isospin breaking.

The primary objective of this work is to substantially improve upon the exploratory calculation of HVP in the functional approach reported in Ref.~\cite{Goecke:2011pe}. That study employed the standard Rainbow-Ladder truncation (RL), whose inherent systematic uncertainties limited the overall precision to the $10\% -20 \%$ level. In this work we employ a setup that goes beyond the standard RL truncation by including explicit pion-exchange and photon-exchange contributions in the quark gap equation and in the inhomogeneous BSE for the quark–photon vertex. In this framework, the $\rho$ resonance and its associated two-pion branch cut emerge dynamically in the timelike quark–photon vertex, leading to a realistic analytic structure that constrains the dressing functions also at low spacelike momenta~\cite{Miramontes:2019mco, Miramontes:2021xgn}. As a result, the spacelike quark–photon vertex entering the HVP integral naturally carries the imprint of the relevant hadronic degrees of freedom rather than through a phenomenological implementation of vector-meson dominance. The present framework reduces the uncertainty of our determination to the few-percent level, representing roughly an order-of-magnitude improvement over the exploratory study of Ref. \cite{Goecke:2011pe}.

Using this approach and formalism, we compute the renormalized scalar vacuum polarization function and obtain the leading-order HVP contribution to the muon anomalous magnetic moment from light, strange, 
charm and bottom quarks, $a_\mu^{\mathrm{HVP,LO}}(u+d+s+c+b)$, in a setup that consistently incorporates strong and electromagnetic isospin breaking at the quark level. The resulting contribution is compared with recent lattice-QCD determinations, which allows us to quantify the net shift induced by isospin breaking, which we find to be a sub-percent
effect. Therefore, our results provide an independent continuum calculation for the hadronic vacuum-polarization contribution, including controlled isospin-breaking corrections and constitute an important benchmark complementary to lattice-QCD calculations.

This paper is organized as follows. In Section~\ref{sec:formalism} we summarize the DSE/BSE framework and introduce the ingredients required for the HVP calculation, in particular the dressed quark propagator, meson BS amplitudes, and the fully dressed quark-photon vertex. Section~\ref{sec:truncation} describes the truncation scheme employed to incorporate pion back-reaction effects together with strong and electromagnetic isospin breaking at the quark level. In Section~\ref{sec:HVP_theory} we briefly review the definition of the hadronic vacuum polarization tensor and the representation used to obtain $a_\mu^{\mathrm{HVP,LO}}$ from the renormalized scalar function $\Pi_R(P^2)$. Our numerical results for light, strange, charm, and bottom-quark contributions to the HVP, including a quantitative assessment of isospin-breaking effects and a comparison with recent lattice-QCD determinations, are presented in Section~\ref{sec:results}. Finally, our conclusions are presented in Section~\ref{sec:conclusions}.

\begin{figure*}[t!]
\centerline{%
\includegraphics[width=1\textwidth]{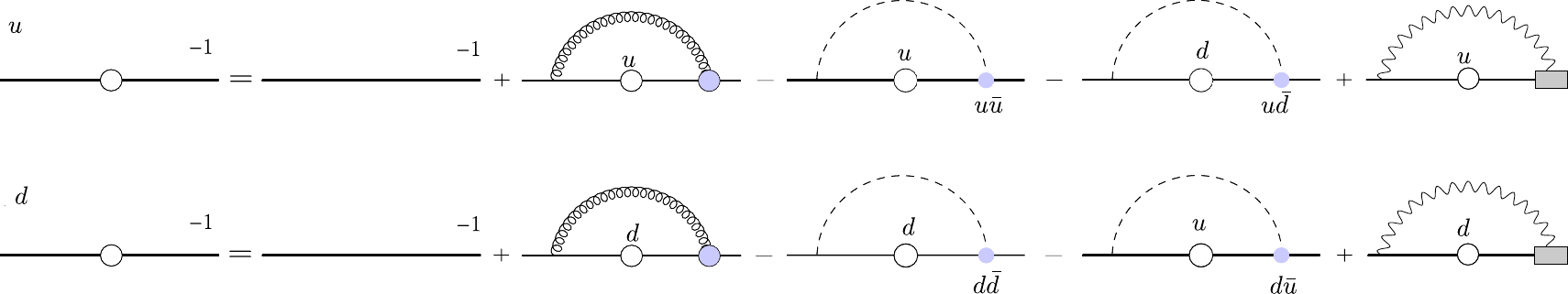}}
\caption{Truncation scheme for the light-quark DSEs. The coupled equations for the $u$ and $d$ quarks include a flavor-blind dressed gluon exchange, explicit pion back-reaction, and a dressed photon loop. Gluon, meson, and photon lines are depicted as curly, dashed, and wavy, respectively. The analogous DSEs for the strange, charm and bottom quarks are not shown but take a similar form, restricted to only the gluon and photon interaction terms.}
\label{fig:truncation}     
\end{figure*}

\section{Dynamical equations}
\label{sec:formalism}

In this section, we outline the main elements of the DSE/BSE approach employed in our analysis of the hadronic vacuum polarization. A detailed discussion of this formalism can be found in Refs.~\cite{Alkofer:2000wg,Eichmann:2016yit,Bashir:2012fs,Sanchis-Alepuz:2017jjd, Eichmann:2025wgs, Huber:2025cbd}. All calculations are performed in Euclidean space.
A central object in this framework is the fully dressed quark propagator $S(p)$, which reflects the nonperturbative phenomenon of dynamical chiral symmetry breaking. This feature is essential for understanding hadron structure and emerges naturally within QCD. For a given quark flavor $f$, the propagator satisfies the corresponding DSE:
\begin{eqnarray} \label{quark-dse}
&& \hspace{-1cm} S^{-1}(p) = Z_2\,(i\slashed{p} + Z_m m_f) \nonumber \\
&& \hspace{-5mm}   - Z_{1f} \,g^2 \,C_F \int \frac{d^4 q}{(2\pi)^4} \, \gamma^\mu \, S(q) \, \Gamma^\nu_{\text{qg}}(q,p) \, D^{\mu \nu}(k) \,,
\end{eqnarray}
where $m_f$ is the current-quark mass, $C_F = 4/3$ is the quadratic Casimir invariant for the fundamental representation of SU(3), and $k = q - p$ is the gluon momentum. The functions $D^{\mu \nu}(k)$ and $\Gamma^\nu_{\text{qg}}(q,p)$ represent the fully dressed gluon propagator and the quark-gluon vertex, respectively. The renormalization constants $Z_2$, $Z_m$, and $Z_{1f}$ are fixed at a chosen renormalization point.

Within the DSE/BSE formalism, mesons emerge naturally and dynamically as relativistic bound states (or resonances) of a quark and an antiquark. Their properties, such as masses and decay constants, 
are encoded in their BS amplitudes (BSA). These 
are solutions of the homogeneous BSE. For example, the BS amplitude $\Gamma_\pi(p,P)$ of the pion satisfies the homogeneous BSE
\begin{equation} \label{eq:homogeneousBSE} 
\begin{split}
[\Gamma_\pi(p,P)]_{\alpha\beta} &= C_F \int \frac{d^4 q}{(2\pi)^4} \, [ \mathbf{K}(p,q,P)]_{\alpha\gamma;\delta\beta} \\
&\quad \times [S(q_+)\, \Gamma_\pi(q,P)\, S(q_-)]_{\gamma\delta} \,,
\end{split}
\end{equation}
where $P$ denotes the total momentum of the bound state, $p$ is the relative momentum between the quark and antiquark, and $q$ is the loop momentum. The internal momenta are defined as $q_\pm = q \pm P/2$, and 
$\alpha,\dots,\delta$ are Dirac indices. The quark propagator $S(q)$ in Eq.~\eqref{eq:homogeneousBSE} is fully dressed and obtained from the solution of the DSE discussed above.

The object $\mathbf{K}(p,q,P)$ is the BSE interaction kernel, which encodes every possible interaction between the quark and antiquark. Its structure depends on the truncation scheme adopted and must be chosen consistently with the quark DSE to preserve the relevant symmetries, such as chiral symmetry. For instance, ensuring that the kernel satisfies the axial-vector Ward-Takahashi identity guarantees that the pion remains massless in the chiral limit. 
The homogeneous BSE, Eq.~\eqref{eq:homogeneousBSE}, determines both the pion Bethe-Salpeter amplitude, $\Gamma_\pi(p,P)$, and the corresponding bound-state mass, $m_\pi$, which is identified through the appearance of a pole in the quark-antiquark scattering matrix at $P^2=-m_\pi^2$. Analogous equations describe other mesons upon choosing the appropriate quantum numbers.

The internal structure of mesons and their interaction with electromagnetic probes are encoded in the dressed quark-photon vertex $\Gamma^\mu$, which plays a pivotal role in describing hadronic structure and processes involving electromagnetic transitions~\cite{Frank:1994mf,Maris:1999bh,Eichmann:2014qva,Miramontes:2019mco,Leutnant:2018dry,Tang:2019zbk}. This vertex describes how a photon couples to a quark in the presence of full QCD dynamics. The quark-photon vertex is determined from the inhomogeneous BSE,
\begin{eqnarray} \label{eq:inhomBSE_vector}
&& \hspace{-4mm} [\Gamma^{\mu}(p,Q)]_{\alpha\beta} =Z_2 \gamma^\mu_{\alpha\beta} + C_F \int \frac{d^4 q}{(2\pi)^4} \, [ \mathbf{K}(p,q,Q)]_{\alpha\gamma;\delta\beta} \nonumber \\
&& \hspace{2cm} \times [S(q_+)\, \Gamma^\mu(q,Q)\, S(q_-)]_{\gamma\delta} \,,
\end{eqnarray}
where $Q$ is the photon momentum and $q_\pm = q \pm Q/2$ are the quark momenta. The kernel $\mathbf{K}$ is the same interaction kernel used in the meson BSE and quark DSE, ensuring consistency across bound-state and vertex calculations.

The quark-photon vertex can be decomposed into a Lorentz-Dirac tensor basis containing twelve linearly independent structures: 
\begin{equation}
\Gamma^\mu(p,Q) = \Gamma^\mu_\text{BC}(p,Q) + \Gamma^\mu_T(p,Q)\,.
\end{equation}
The longitudinal part $\Gamma^\mu_\text{BC}$ is given by the Ball-Chiu construction~\cite{Ball:1980ay} and is completely fixed by the quark propagator through the Ward-Takahashi identity. It ensures current conservation and encodes essential nonperturbative dressing effects. The transverse part $\Gamma^\mu_T$, orthogonal to $Q^\mu$, contains eight additional tensor structures that are unconstrained by the Ward identity but are crucial for capturing resonance phenomena.

In our study, we retain all twelve components of the vertex to maintain a complete and self-consistent description. The transverse part, in particular, dynamically generates vector-meson poles such as the $\rho$ resonance including a finite width due to its decay to two pions \cite{Williams:2018adr, Miramontes:2021xgn}.

\section{Truncation Scheme}
\label{sec:truncation}

Functional approaches based on the DSE have traditionally relied on the RL truncation. In this framework, the product of the dressed gluon propagator and the dressed quark–gluon vertex entering the quark gap equation is replaced by an effective interaction. Its ultraviolet behaviour is constrained by perturbative QCD, while its infrared strength is fixed by requiring a simultaneous description of key hadronic observables, most notably the pion mass and decay constant. In the Landau gauge, the gluon propagator is defined as:
\begin{equation}
D^{\mu\nu}(k)=T^{\mu\nu}(k)\frac{Z(k^2)}{k^2} \,,
\end{equation}
where $T^{\mu\nu}$ is the transverse projector and  $Z(k^2)$ is the gluon dressing function. Within the RL truncation, the product of the dressed gluon propagator and the approximate quark-gluon vertex, $ \Gamma^{\nu}_{\rm qg}(q, p) \approx \gamma^{\nu} \Gamma(k^2)$, is absorbed into an effective running interaction according to
\begin{align}
Z_{1f} \,\frac{g^2}{4 \pi}\, Z(k^2) \, \Gamma(k^2) \rightarrow Z^2_2 \,  \alpha_{\mathrm{eff}}(k^2)\,.
\end{align}

To consistently incorporate both strong and electromagnetic effects at the quark level, we employ a truncation scheme that extends the conventional RL approximation by including explicit meson- and photon-exchange contributions to the quark self-energy; see, e.g., Refs.~\cite{Miramontes:2021xgn,Miramontes:2022uyi,Miramontes:2022mex}. Consequently, all Green's functions entering the calculation are obtained from a coupled system of 
DSEs/BSEs solved in an isospin-broken framework.

The dressed quark propagators are determined from the quark DSEs. Whereas the standard RL truncation retains only the flavor-blind dressed gluon exchange, the present scheme additionally incorporates meson back-reaction through effective meson-exchange loops and dressed photon exchange. These corrections provide the leading strong and electromagnetic contributions beyond RL while preserving the self-consistent DSE/BSE framework. The corresponding truncation of the quark DSE is illustrated in Fig.~\ref{fig:truncation}.

The gluon-mediated interaction is modeled by an effective running coupling of Maris-Tandy (MT) type \cite{Maris:1997hd},
\begin{align}
\alpha_{\mathrm{eff}}(k^2)
&=
\pi \eta^7\,
\left(\frac{k^2}{\Lambda^2}\right)^2
e^{\!-\eta^2\frac{k^2}{\Lambda^2}}
\nonumber\\
&\quad+
\frac{2\pi \gamma_m (1 - e^{\!-k^2/4\Lambda_t^2})}{\ln\!\left[
e^2 - 1 + \left(1 + k^2/\Lambda_{\mathrm{QCD}}^2\right)^2
\right]}
\,,
\end{align}
which reproduces dynamical chiral symmetry breaking in the infrared and the correct one-loop ultraviolet behavior. In this effective interaction, the scale $\Lambda_t = 1\,\mathrm{GeV}$ is introduced for technical convenience. The anomalous dimension is fixed to $\gamma_m = 12/(11N_c - 2N_f) = 12/25$, corresponding to $N_f = 4$ active quark flavors and $N_c = 3$ colors. The QCD scale is set to $\Lambda_{\mathrm{QCD}} = 0.234\,\mathrm{GeV}$. The remaining infrared parameters $\eta$ and $\Lambda$ are discussed below.

\subsection{Strong Isospin Breaking}
\label{subsec:strong_ib}

Strong isospin breaking is incorporated through the explicit assignment of distinct current masses to the up and down quarks in the quark DSEs. As a consequence, the resulting dressed quark propagators differ for the two light flavors.

In an RL-only setup the BSEs for $u\bar{u}$ and $d\bar{d}$ states are fully decoupled.  However, once meson-exchange contributions are included in the interaction kernel, these two channels become coupled, as illustrated in Figure~\ref{fig:truncation}. This coupling is a direct consequence of pion exchange and is physically necessary: the neutral pion is a mixed $u\bar{u}$/$d\bar{d}$ state that can only be described correctly if the two amplitudes are solved simultaneously, with the coupling generated dynamically by the meson-exchange kernel and reflecting the explicit breaking of isospin symmetry at the quark level.

\begin{figure}[t!]
\centerline{%
\includegraphics[width=0.35\textwidth]{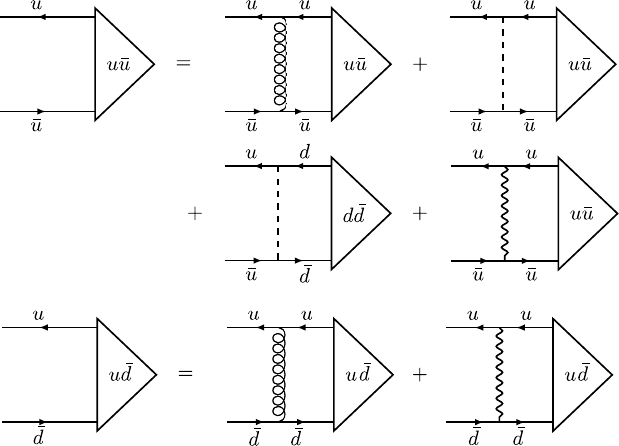}}
\caption{Truncation for the coupled system of BSE in the neutral and charged pseudoscalar channels. An equation for the $d\bar d$ amplitude corresponding to the $u\bar u$ amplitude is obtained by interchanging $u\bar u \leftrightarrow d\bar d$.}
\label{fig:truncation_BSA}     
\end{figure}

In contrast, the charged channels $u\bar d$ and $d\bar u$ do not receive contributions from meson exchange within the present truncation and are therefore described by BSEs containing only RL gluon exchange and photon exchange. As a result, strong isospin breaking in these channels enters solely through the explicit quark-mass dependence of the propagators.

\begin{figure}[t!]
\centerline{%
\includegraphics[width=0.3\textwidth]{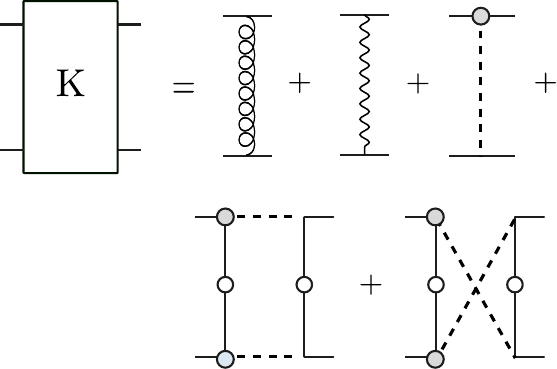}}
\caption{Truncation of the quark-quark scattering kernel. Curly, wavy, dashed and solid lines denote gluon, photon, meson and quark propagators, respectively.}
\label{fig:truncation_qphv}     
\end{figure}

The meson-exchange contributions are constructed following an axial-vector Ward-Takahashi identity preserving scheme and are dominated by pion exchange in the light-quark sector \cite{Fischer:2008wy,Fischer:2008sp, Sanchis-Alepuz:2014wea}.
The corresponding integral representations for Figures~\ref{fig:truncation} and \ref{fig:truncation_BSA} can be found in \cite{Miramontes:2022mex} while the expression for Figure~\ref{fig:truncation_qphv} in \cite{Miramontes:2019mco}.
Contributions from heavier mesons are suppressed by their larger masses and are neglected in the present work. In this way, strong isospin breaking arises from the interplay between explicit current quark-mass differences and nonperturbative quark-antiquark correlations beyond the RL approximation.

\subsection{Electromagnetic Isospin Breaking}
\label{subsec:em_ib}

Electromagnetic isospin breaking is incorporated by including dressed photon exchange contributions in all dynamical equations of the truncation. Photon exchange enters the quark DSE as an additional self-energy term and appears consistently in the interaction kernels of both the homogeneous and inhomogeneous BSEs, see Figures~\ref{fig:truncation} and \ref{fig:truncation_BSA}.

The quark-photon vertex $\Gamma_\mu$ entering physical amplitudes is obtained from the solution of the inhomogeneous BSE, Eq.~(\ref{eq:inhomBSE_vector}), and is treated in its full tensor decomposition, retaining all twelve Dirac structures and their associated dressing functions. This provides a complete description of the quark-photon interaction. This full vertex is also used in the HVP quark-loop diagram, see Eq.~(\ref{eq:hvp}) below.

In contrast, inside the photon-exchange kernel it is sufficient to only take the leading $\gamma^\mu$ component of the vertex into account, since the electromagnetic coupling $\alpha_0 \simeq 1/137$ makes any refinement of its tensor structure a negligible higher-order effect in $\alpha_{\rm QED}$.
In this case, the QED contribution takes the same form as the gluon RL term and can be incorporated by adding
\begin{equation}
\alpha_{\mathrm{QED}}(k^2)
=
\alpha_0\, Q_f\, Q_g\, Z_{\mathrm{QED}}(k^2)
\end{equation}
to the effective strong interaction. Here $Q_f$, $Q_g$ are the electric charges of the two quarks attached to the photon line. Moreover, the dressing of the leading tensor structure of the quark-photon vertex is uniquely determined by a Ward-identity and given by 
\begin{equation}
Z_{\mathrm{QED}}(k^2)
=
\frac{A_f(k^2)+A_g(k^2)}{2},
\end{equation}
where $A_f$ and $A_g$ are the dressing functions of the corresponding quark propagators.

\section{Leading order HVP}
\label{sec:HVP_theory}

The leading-order hadronic contribution to the anomalous magnetic moment of the muon, $ a_\mu^{\text{HVP}}$, arises from vacuum polarization effects due to the strong interactions. These contributions are encapsulated by the renormalized hadronic vacuum polarization scalar function $\Pi_R(P^2) $. 
The expression for $a_\mu^{\text{HVP}}$ is given by \cite{Greynat:2022geu}:
\begin{equation}
a_\mu^{\text{HVP}} = \frac{\alpha_0}{\pi} \int_0^1 dx \, (1 - x) \left[ -e^2 \Pi_R\left( \frac{x^2}{1 - x} m_\mu^2 \right) \right],
\label{eq:hvp}
\end{equation}
where $m_\mu$ denotes the muon mass, and $\Pi_R(P^2)$ is the subtracted vacuum polarization function, renormalized such that $\Pi_R(0)=0$ to preserve charge non-renormalization. 

The photon vacuum polarization tensor is computed from non-perturbative QCD input and corresponds to the self-energy of the photon DSE. The tensor is written as follows,

\begin{equation}
\Pi_{\mu\nu}(P) = Z_2 \int \frac{d^4 q}{(2\pi)^4} \, \text{Tr} \left[ S(q_-) \, \Gamma_\mu(q, P) \, S(q_+) \, \gamma_\nu \right].
\label{eq:hvp_tensor}
\end{equation}

To extract the scalar vacuum polarization function $\Pi(P^2)$, we project the tensor onto its transverse component, using the Lorentz structure dictated by gauge invariance\footnote{Note that this can be accomplished by contracting $\Pi_{\mu\nu}(P)$ with a tensor of the form  
		$\left( \delta_{\mu\nu} - \zeta \frac{P_\mu P_\nu}{P^2} \right)$ and arbitrary values of $\zeta$. This provides 
		a natural way to assess the numerical error of our calculations: due to inaccuracies, the vacuum polarization tensor 
		may contain spurious longitudinal components, whose impact can be addressed by varying $\zeta$. Brown and Pennington 
		suggested in particular comparing differences in results for $\zeta=1$ and $\zeta=4$, see \cite{Brown:1988bn} for details. We adopt this 
		procedure also in this work. }:

\begin{equation}
\Pi_{\mu\nu}(P) = \left( \delta_{\mu\nu} - \frac{P_\mu P_\nu}{P^2} \right) P^2 \Pi(P^2).
\label{eq:projection}
\end{equation}

The scalar function $\Pi(P^2)$ is UV divergent. We remove the quadratic divergence through
\begin{equation}
[P^2\Pi(P^2)]_R := P^2\Pi(P^2) - \epsilon^2\Pi(\epsilon^2),
\end{equation}
with $\epsilon$ chosen as small as numerically possible. The remaining logarithmic divergence is fixed by matching the infrared behaviour to $\Pi(P^2)\propto P^2$. Writing the renormalized function as 
\begin{equation}
\Pi_R(P^2) := \Pi(P^2) - \Pi(P_0^2) + a\,P^2,
\label{eq:hvp_renorm}
\end{equation}
with $P_0^2$ a matching point in that region.

The remaining ingredients that define the vacuum polarization function are the quark propagator and the quark-photon vertex. In this work, we employ the truncation depicted in Fig.~\ref{fig:truncation}. 
Once the quark DSE and the quark-photon vertex have been computed by standard techniques, Eq.~\eqref{eq:hvp_tensor} can be calculated for different quark flavors. From here, the scalar function Eq.~\eqref{eq:hvp_renorm} can be extracted and inserted into Eq.~\eqref{eq:hvp} to compute the contribution to the $g-2$ of the muon.

\section{Numerical Results}
\label{sec:results}

\begin{figure*}[t!]
\centerline{%
\includegraphics[width=0.80\textwidth]{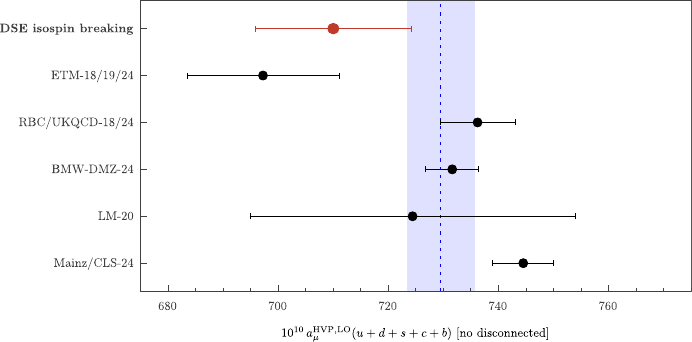}}
\caption{Our result for $a_{\mu}^{\mathrm{HVP,LO}}$ including strong and electromagnetic isospin breaking, compared with different lattice calculations for $u+d+s+c+b$ flavors. The results of this work are compared to those found by the ETM-18/19/24~\cite{Giusti:2018mdh,Giusti:2019hkz,ExtendedTwistedMass:2024nyi}, RBC/UKQCD-18/24~\cite{RBC:2018dos,RBC:2024fic}, BMW-DMZ-24~\cite{Boccaletti:2024guq}, LM-20~\cite{Lehner:2020crt} and Mainz/CLS-24~\cite{Djukanovic:2024cmq} lattice collaborations (without disconnected contributions). The blue band stands for the Muon g-2 Theory Initiative recommended value~\cite{Aliberti:2025beg}, resulting from the combination of the lattice computations shown.}
\label{fig:HVP_result}     
\end{figure*}

First, we specify the QCD input, namely the current quark masses $m_{u}$, $m_{d}$ and $m_s$, the heavy-quark masses $m_c$ and $m_b$, and the effective interaction characterized by the parameters $\Lambda$ and $\eta$. We fix the interaction scale to $\Lambda = 0.8\,\mathrm{GeV}$ and choose $\eta = 1.45 \pm 0.10$. This choice yields a reasonable simultaneous description of the pion and $\rho$-meson masses, with the $\rho^\pm$ mass given by $m_\rho = 766.3\,\mathrm{MeV}$ and varying by approximately $\pm 9\,\mathrm{MeV}$ over the quoted range of $\eta$. The resulting spread provides an estimate of the associated model uncertainty.

With the interaction parameters fixed, the up, down, and strange quark masses are then determined by requiring that a calculation including RL gluon exchange, meson exchange, and photon exchange reproduces well the experimental masses of the charged pion and of the neutral and charged kaons. This procedure yields $m_u = 5.7\,\mathrm{MeV}$, $m_d = 7.2\,\mathrm{MeV}$, and $m_s = 81\,\mathrm{MeV}$ for the quark masses, defined in a MOM renormalization scheme at a renormalization scale $\mu = 19\,\mathrm{GeV}$. We emphasize that these values are input parameters fixed by the above fitting procedure and should therefore not be regarded as predictions. The charm- and bottom-quark masses are determined analogously, by requiring agreement with the experimental masses of the $J/\Psi$ and the $\Upsilon(1S)$. This yields
$m_c = 830\,\mathrm{MeV}$ and $m_b = 3690\,\mathrm{MeV}$. 

Before discussing the hadronic vacuum polarization, we briefly summarize the light-meson spectrum produced by the present truncation. The inclusion of strong and electromagnetic isospin breaking lifts the degeneracy within the pseudoscalar and vector meson multiplets through the combined effects of unequal light-quark masses and electromagnetic interactions. Table \ref{tab:meson_masses} lists the masses of the $\pi^0$, $\pi^\pm$, $K^0$, $K^\pm$, $\rho^0$, and $\rho^\pm$ states obtained from the corresponding homogeneous BSEs, together with their experimental values.

For the parameter choice discussed above, the corresponding decay constants are slightly enhanced. In the pseudoscalar sector we obtain
$f_\pi = 135(1)\,\mathrm{MeV}$ and $f_K = 157(2)\,\mathrm{MeV}$, to be compared with the experimental values $f_\pi^{\mathrm{exp}} = 130.2(1.2)\,\mathrm{MeV}$ and $f_K^{\mathrm{exp}} = 155.7(3)\,\mathrm{MeV}$. In the vector channel, the $\rho$-meson decay constant comes out as $f_\rho = 226(2)$ MeV, close to the value $f_\rho^{\mathrm{exp}} = 220$ MeV.
Such deviations at the few-percent level are typical for extended truncations beyond RL \cite{Miramontes:2021xgn,Miramontes:2021exi, Xu:2022kng} and reflect the
different sensitivity of masses and decay constants to details of the interaction kernel.

\begin{table}[H]
\centering
\caption{Light meson masses obtained in the present truncation including strong and electromagnetic isospin breaking, compared to experimental values \cite{ParticleDataGroup:2026aaa}. All masses are given in MeV.}
\label{tab:meson_masses}
\begin{tabular}{lcc}
\hline\hline
State & This work & PDG \\
\hline
$\pi^0$        & 135.2 & 135.0
\\
$\pi^\pm$      & 140.8 & 139.
6\\
$K^0$          & 497.1 & 497.
6\\
$K^\pm$        & 493.6 & 493.7 \\
$\rho^0$       & 767.0 & 775.
3(2)\\
$\rho^\pm$     & 766.3 & 775.1(3)\\
\hline\hline
\end{tabular}
\end{table}

\subsection{HVP result}
\label{subsect:HVP_results}

We compute the leading-order hadronic vacuum polarization contribution to the anomalous magnetic moment of the muon from the renormalized scalar function $\Pi_R(P^2)$ introduced in Sec.~\ref{sec:HVP_theory}. The calculation employs the dynamically dressed quark-photon vertex and quark propagators obtained within the truncation scheme detailed in Sec.~\ref{sec:truncation}. The resulting scalar function is then inserted into Eq.~\eqref{eq:hvp} to obtain the contribution of each quark flavor to $a_\mu^{\rm{HVP}}$.

The overall uncertainty receives contributions from three identifiable sources. The dominant one originates from the dependence on the parameter $\eta$ in the MT interaction and amounts to approximately 2\% of the total result. The uncertainty associated with the renormalization of the vacuum polarization tensor is below 0.1\%. Finally, comparing the transverse projection of Eq.~\eqref{eq:projection} with the Brown-Pennington projection (see the footnote above Eq.~\eqref{eq:projection}) yields a difference of 0.5\%, which we attribute entirely to numerical effects.

Including strong and electromagnetic isospin breaking (ISB), we obtain
for the flavor content most commonly quoted by lattice collaborations
\begin{equation}
  a_\mu^{\text{HVP,LO}}(u+d+s+c)\big|_{\text{ISB}}
  = 709.7 \times 10^{-10}\,.
  \label{eq:hvp4f}
\end{equation}
The bottom-quark contribution is strongly suppressed by its large
mass; we find $a_\mu^{\text{HVP,LO}}(b) = 0.3 \times 10^{-10}$. Including it, and adding the uncertainty sources discussed above in
quadrature, we obtain our final result
\begin{equation}
  a_\mu^{\text{HVP,LO}}(u+d+s+c+b)\big|_{\text{ISB}}
  = (710.0 \pm 14.5) \times 10^{-10}\,.
  \label{eq:hvpfinal}
\end{equation}

Figure~\ref{fig:HVP_result} compares this total result with recent lattice-QCD determinations, whose disconnected contributions were subtracted~\footnote{For the BMW-DMZ result~\cite{Boccaletti:2024guq}, we subtracted the disconnected contributions quoted as ``Avg.2B" in the second White Paper of the Muon g-2 Theory Initiative (WP25), $a_\mu^{\rm{HVP,LO}}\Big|^{\rm lat}_{\rm Avg.2B}(\rm disc)=-16.5(2.3)\times10^{-10}$~\cite{Aliberti:2025beg}. The blue band in Figure~\ref{fig:HVP_result} was obtained by applying the same subtraction to the consolidated lattice average quoted in the WP25, $a_\mu^{\rm{HVP,LO}}\Big|^{\rm lat}_{\rm WP 25}=713.2(6.1)\times10^{-10}$~\cite{Aliberti:2025beg}.}.

Within the quoted uncertainties, the DSE/BSE result remains compatible with these determinations. Moreover, it lies close to the global average obtained from their combination. This level of compatibility indicates that our truncation, which incorporates pion back-reaction effects together with strong and electromagnetic isospin breaking at the quark level, captures the essential non-perturbative dynamics of the HVP.

Finally, to quantify the impact of isospin breaking, we repeat the calculation in the isospin-symmetric (IS) limit by setting $m_d = m_u$ and $\alpha_0=0$. This yields
\begin{equation}
  a_\mu^{\mathrm{HVP,LO}}(ud+s+c+b)\big|_{\mathrm{IS}}
  = 705.5 \times 10^{-10}\,.
\end{equation}
Comparing this result with the full isospin-breaking (ISB) calculation gives
\begin{equation}
  \begin{aligned}
    \Delta a_\mu^{\mathrm{HVP,LO}}
    &\equiv a_\mu^{\mathrm{HVP,LO}}\big|_{\mathrm{ISB}}
      -  a_\mu^{\mathrm{HVP,LO}}\big|_{\mathrm{IS}} \\
    &= 4.5 \times 10^{-10}\,,
  \end{aligned}
\end{equation}
corresponding to a relative change of about 0.6\%. Although quantitatively modest, this result demonstrates that strong and electromagnetic isospin breaking produces a non-negligible correction within the present DSE/BSE framework.

\section{Conclusions}
\label{sec:conclusions}
In this work, we have presented a determination of the leading-order hadronic vacuum polarization contribution to the muon anomalous magnetic moment,
 $a_\mu^{\mathrm{HVP,LO}}(u+d+s+c+b)$ 
within a functional approach to QCD based on the coupled Dyson-Schwinger and Bethe-Salpeter equations. The present calculation represents a significant advance over previous functional studies through a substantially improved truncation scheme. In particular, it consistently incorporates meson back-reaction effects in the quark self-energy, a fully dressed quark-photon vertex featuring a dynamical $\rho$-meson resonance with finite-width effects, and both strong and electromagnetic isospin breaking at the quark level.

These developments considerably enhance the predictive power of the approach. Most notably, they remove the ambiguity in the determination of the light-quark masses that constituted the dominant source of uncertainty in earlier functional calculations \cite{Goecke:2011pe}, leading instead to an improved determination of the light-quark sector together with a substantially reduced theoretical uncertainty. Our final result, see Eq.~(\ref{eq:hvpfinal}),
is in good agreement with recent lattice-QCD determinations for the same quark-flavor content. Quantitatively, isospin breaking shifts the isospin-symmetric result by 0.6\%. 

The present framework provides a unified and self-consistent description of the nonperturbative strong and electromagnetic dynamics relevant to the hadronic vacuum polarization. Since all Green's functions are obtained within the same coupled DSE/BSE framework, the approach can be systematically improved by refining the underlying truncation while preserving the symmetries of QCD. Future extensions will include higher-order hadronic contributions, a more complete treatment of resonance dynamics, and further improvements of the quark-gluon interaction.

\section*{Acknowledgements}
 A.\,S.\,M. is funded by the Spanish MICINN grants PID2020-113334GB-I00 and PID2023-151418NB-I00, the Generalitat Valenciana grant CIPROM/2022/66, and CEX2023-001292-S by MCIU/AEI.
 P.\,R. thanks the hospitality and financial support of IFIC and Universitat de Val\`encia, where part of this work was done. A.\,B. acknowledges the Beatriz-Galindo support at the University of Huelva, Huelva, Spain. P.\,R. was partly funded by Conahcyt-Secihti (Mexico) through project CBF2023-2024-3226, and also MCIN/AEI/10.13039/501100011033 (Spain), grants PID2020-114473GB-I00, and  PID2023-146220NB-I00, and by Generalitat Valenciana (Spain), grants PROMETEO2021/071 and CIESGT-2024-21.
\bibliography{main}

@article{Eichmann:2016yit,
    author = "Eichmann, Gernot and Sanchis-Alepuz, Helios and Williams, Richard and Alkofer, Reinhard and Fischer, Christian S.",
    title = "{Baryons as relativistic three-quark bound states}",
    eprint = "1606.09602",
    archivePrefix = "arXiv",
    primaryClass = "hep-ph",
    doi = "10.1016/j.ppnp.2016.07.001",
    journal = "Prog. Part. Nucl. Phys.",
    volume = "91",
    pages = "1--100",
    year = "2016"
}

@article{Bashir:2012fs,
    author = "Bashir, Adnan and Chang, Lei and Cloet, Ian C. and El-Bennich, Bruno and Liu, Yu-Xin and Roberts, Craig D. and Tandy, Peter C.",
    title = "{Collective perspective on advances in Dyson-Schwinger Equation QCD}",
    eprint = "1201.3366",
    archivePrefix = "arXiv",
    primaryClass = "nucl-th",
    doi = "10.1088/0253-6102/58/1/16",
    journal = "Commun. Theor. Phys.",
    volume = "58",
    pages = "79--134",
    year = "2012"
}

@article{Raya:2016yuj,
    author = "Raya, Khepani and Ding, Minghui and Bashir, Adnan and Chang, Lei and Roberts, Craig D.",
    title = "{Partonic structure of neutral pseudoscalars via two photon transition form factors}",
    eprint = "1610.06575",
    archivePrefix = "arXiv",
    primaryClass = "nucl-th",
    doi = "10.1103/PhysRevD.95.074014",
    journal = "Phys. Rev. D",
    volume = "95",
    number = "7",
    pages = "074014",
    year = "2017"
}

@article{Miramontes:2022uyi,
    author = "Miramontes, A. S. and Bashir, Adnan",
    title = "{Timelike electromagnetic kaon form factor}",
    eprint = "2212.10800",
    archivePrefix = "arXiv",
    primaryClass = "hep-ph",
    doi = "10.1103/PhysRevD.107.014016",
    journal = "Phys. Rev. D",
    volume = "107",
    number = "1",
    pages = "014016",
    year = "2023"
}

@article{Ding:2018xwy,
    author = "Ding, Minghui and Raya, Khepani and Bashir, Adnan and Binosi, Daniele and Chang, Lei and Chen, Muyang and Roberts, Craig D.",
    title = "{$\gamma^\ast \gamma \to \eta, \eta^\prime$ transition form factors}",
    eprint = "1810.12313",
    archivePrefix = "arXiv",
    primaryClass = "nucl-th",
    doi = "10.1103/PhysRevD.99.014014",
    journal = "Phys. Rev. D",
    volume = "99",
    number = "1",
    pages = "014014",
    year = "2019"
}

@article{Sanchis-Alepuz:2017jjd,
    author = "Sanchis-Alepuz, Helios and Williams, Richard",
    title = "{Recent developments in bound-state calculations using the Dyson\textendash{}Schwinger and Bethe\textendash{}Salpeter equations}",
    eprint = "1710.04903",
    archivePrefix = "arXiv",
    primaryClass = "hep-ph",
    doi = "10.1016/j.cpc.2018.05.020",
    journal = "Comput. Phys. Commun.",
    volume = "232",
    pages = "1--21",
    year = "2018"
}

@article{Maris:1997hd,
    author = "Maris, Pieter and Roberts, Craig D. and Tandy, Peter C.",
    title = "{Pion mass and decay constant}",
    eprint = "nucl-th/9707003",
    archivePrefix = "arXiv",
    reportNumber = "ANL-PHY-8753-TH-97, KSUCNR-103-97",
    doi = "10.1016/S0370-2693(97)01535-9",
    journal = "Phys. Lett. B",
    volume = "420",
    pages = "267--273",
    year = "1998"
}

@article{Miramontes:2019mco,
    author = "Miramontes, \'Angel S. and Sanchis-Alepuz, H\`elios",
    title = "{On the effect of resonances in the quark-photon vertex}",
    eprint = "1906.06227",
    archivePrefix = "arXiv",
    primaryClass = "hep-ph",
    doi = "10.1140/epja/i2019-12847-6",
    journal = "Eur. Phys. J. A",
    volume = "55",
    number = "10",
    pages = "170",
    year = "2019"
}

@article{Eichmann:2014qva,
    author = "Eichmann, Gernot",
    editor = "Bicudo, Pedro and Giacosa, Francesco and Malek, Magdalena and Marinkovic, Marina and Parganlija, Denis",
    title = "{Probing nucleons with photons at the quark level}",
    eprint = "1404.4149",
    archivePrefix = "arXiv",
    primaryClass = "nucl-th",
    doi = "10.5506/APhysPolBSupp.7.597",
    journal = "Acta Phys. Polon. Supp.",
    volume = "7",
    number = "3",
    pages = "597",
    year = "2014"
}

@article{Maris:1999bh,
    author = "Maris, Pieter and Tandy, Peter C.",
    title = "{The Quark photon vertex and the pion charge radius}",
    eprint = "nucl-th/9910033",
    archivePrefix = "arXiv",
    reportNumber = "KSUCNR-112-99",
    doi = "10.1103/PhysRevC.61.045202",
    journal = "Phys. Rev. C",
    volume = "61",
    pages = "045202",
    year = "2000"
}

@article{Leutnant:2018dry,
    author = "Leutnant, Milena and Sternbeck, Andr\'e",
    title = "{Quark-photon vertex from lattice QCD in Landau gauge}",
    eprint = "1812.11131",
    archivePrefix = "arXiv",
    primaryClass = "hep-lat",
    doi = "10.22323/1.336.0095",
    journal = "PoS",
    volume = "Confinement2018",
    pages = "095",
    year = "2018"
}

@article{Tang:2019zbk,
    author = "Tang, Can and Gao, Fei and Liu, Yu-Xin",
    title = "{Practical scheme from QCD to phenomena via Dyson-Schwinger equations}",
    eprint = "1902.01679",
    archivePrefix = "arXiv",
    primaryClass = "hep-ph",
    doi = "10.1103/PhysRevD.100.056001",
    journal = "Phys. Rev. D",
    volume = "100",
    number = "5",
    pages = "056001",
    year = "2019"
}

@article{Frank:1994mf,
    author = "Frank, M. R.",
    title = "{Nonperturbative aspects of the quark - photon vertex}",
    eprint = "nucl-th/9403009",
    archivePrefix = "arXiv",
    reportNumber = "DOE-ER-40561-131, INT-94-00-52",
    doi = "10.1103/PhysRevC.51.987",
    journal = "Phys. Rev. C",
    volume = "51",
    pages = "987--998",
    year = "1995"
}

@article{Miramontes:2021xgn,
    author = "Miramontes, \'Angel S. and Sanchis Alepuz, H\`elios and Alkofer, Reinhard",
    title = "{Elucidating the effect of intermediate resonances in the quark interaction kernel on the timelike electromagnetic pion form factor}",
    eprint = "2102.12541",
    archivePrefix = "arXiv",
    primaryClass = "hep-ph",
    doi = "10.1103/PhysRevD.103.116006",
    journal = "Phys. Rev. D",
    volume = "103",
    number = "11",
    pages = "116006",
    year = "2021"
}

@article{Miramontes:2021exi,
    author = "Miramontes, \'Angel and Bashir, Adnan and Raya, Kh\'epani and Roig, Pablo",
    title = "{Pion and Kaon box contribution to a\ensuremath{\mu}HLbL}",
    eprint = "2112.13916",
    archivePrefix = "arXiv",
    primaryClass = "hep-ph",
    doi = "10.1103/PhysRevD.105.074013",
    journal = "Phys. Rev. D",
    volume = "105",
    number = "7",
    pages = "074013",
    year = "2022"
}

@article{Raya:2015gva,
    author = "Raya, K. and Chang, L. and Bashir, A. and Cobos-Martinez, J. J. and Guti\'errez-Guerrero, L. X. and Roberts, C. D. and Tandy, P. C.",
    title = "{Structure of the neutral pion and its electromagnetic transition form factor}",
    eprint = "1510.02799",
    archivePrefix = "arXiv",
    primaryClass = "nucl-th",
    doi = "10.1103/PhysRevD.93.074017",
    journal = "Phys. Rev. D",
    volume = "93",
    number = "7",
    pages = "074017",
    year = "2016"
}

@article{Eichmann:2019bqf,
    author = "Eichmann, Gernot and Fischer, Christian S. and Williams, Richard",
    title = "{Kaon-box contribution to the anomalous magnetic moment of the muon}",
    eprint = "1910.06795",
    archivePrefix = "arXiv",
    primaryClass = "hep-ph",
    doi = "10.1103/PhysRevD.101.054015",
    journal = "Phys. Rev. D",
    volume = "101",
    number = "5",
    pages = "054015",
    year = "2020"
}

@article{Eichmann:2019tjk,
    author = "Eichmann, Gernot and Fischer, Christian S. and Weil, Esther and Williams, Richard",
    title = "{Single pseudoscalar meson pole and pion box contributions to the anomalous magnetic moment of the muon}",
    eprint = "1903.10844",
    archivePrefix = "arXiv",
    primaryClass = "hep-ph",
    doi = "10.1016/j.physletb.2019.134855",
    journal = "Phys. Lett. B",
    volume = "797",
    pages = "134855",
    year = "2019",
    note = "[Erratum: Phys.Lett.B 799, 135029 (2019)]"
}

@article{Sanchis-Alepuz:2013iia,
    author = "Sanchis-Alepuz, Helios and Williams, Richard and Alkofer, Reinhard",
    title = "{Delta and Omega electromagnetic form factors in a three-body covariant Bethe-Salpeter approach}",
    eprint = "1302.6048",
    archivePrefix = "arXiv",
    primaryClass = "hep-ph",
    doi = "10.1103/PhysRevD.87.096015",
    journal = "Phys. Rev. D",
    volume = "87",
    number = "9",
    pages = "096015",
    year = "2013"
}

@article{Williams:2018adr,
    author = "Williams, Richard",
    title = "{Vector mesons as dynamical resonances in the Bethe\textendash{}Salpeter framework}",
    eprint = "1804.11161",
    archivePrefix = "arXiv",
    primaryClass = "hep-ph",
    doi = "10.1016/j.physletb.2019.134943",
    journal = "Phys. Lett. B",
    volume = "798",
    pages = "134943",
    year = "2019"
}

@article{Miramontes:2025ofw,
    author = "Miramontes, Angel S. and Eichmann, Gernot and Alkofer, Reinhard",
    title = "{Timelike form factor for the anomalous process {\ensuremath{\gamma}}{\textasteriskcentered}{\ensuremath{\pi}} {\textrightarrow} {\ensuremath{\pi}}{\ensuremath{\pi}}}",
    eprint = "2504.20899",
    archivePrefix = "arXiv",
    primaryClass = "hep-ph",
    doi = "10.1016/j.physletb.2025.139659",
    journal = "Phys. Lett. B",
    volume = "868",
    pages = "139659",
    year = "2025"
}

@article{Stamen:2022uqh,
    author = "Stamen, Dominik and Hariharan, Deepti and Hoferichter, Martin and Kubis, Bastian and Stoffer, Peter",
    title = "{Kaon electromagnetic form factors in dispersion theory}",
    eprint = "2202.11106",
    archivePrefix = "arXiv",
    primaryClass = "hep-ph",
    reportNumber = "PSI-PR-22-04, ZU-TH 22/22",
    doi = "10.1140/epjc/s10052-022-10348-3",
    journal = "Eur. Phys. J. C",
    volume = "82",
    number = "5",
    pages = "432",
    year = "2022"
}

@article{Giusti:2018mdh,
    author = "Giusti, D. and Sanfilippo, F. and Simula, S.",
    title = "{Light-quark contribution to the leading hadronic vacuum polarization term of the muon $g-2$ from twisted-mass fermions}",
    eprint = "1808.00887",
    archivePrefix = "arXiv",
    primaryClass = "hep-lat",
    doi = "10.1103/PhysRevD.98.114504",
    journal = "Phys. Rev. D",
    volume = "98",
    number = "11",
    pages = "114504",
    year = "2018"
}

@article{Giusti:2019hkz,
    author = "Giusti, D. and Simula, S.",
    title = "{Lepton anomalous magnetic moments in Lattice QCD+QED}",
    eprint = "1910.03874",
    archivePrefix = "arXiv",
    primaryClass = "hep-lat",
    doi = "10.22323/1.363.0104",
    journal = "PoS",
    volume = "LATTICE2019",
    pages = "104",
    year = "2019"
}

@article{ExtendedTwistedMass:2024nyi,
    author = "Alexandrou, C. and others",
    collaboration = "Extended Twisted Mass",
    title = "{Strange and charm quark contributions to the muon anomalous magnetic moment in lattice QCD with twisted-mass fermions}",
    eprint = "2411.08852",
    archivePrefix = "arXiv",
    primaryClass = "hep-lat",
    reportNumber = "CERN-TH-2024-197",
    doi = "10.1103/PhysRevD.111.054502",
    journal = "Phys. Rev. D",
    volume = "111",
    number = "5",
    pages = "054502",
    year = "2025"
}

@article{Borsanyi:2020mff,
    author = "Borsanyi, Sz. and others",
    title = "{Leading hadronic contribution to the muon magnetic moment from lattice QCD}",
    eprint = "2002.12347",
    archivePrefix = "arXiv",
    primaryClass = "hep-lat",
    doi = "10.1038/s41586-021-03418-1",
    journal = "Nature",
    volume = "593",
    number = "7857",
    pages = "51--55",
    year = "2021"
}

@article{Lehner:2020crt,
    author = "Lehner, Christoph and Meyer, Aaron S.",
    title = "{Consistency of hadronic vacuum polarization between lattice QCD and the R-ratio}",
    eprint = "2003.04177",
    archivePrefix = "arXiv",
    primaryClass = "hep-lat",
    doi = "10.1103/PhysRevD.101.074515",
    journal = "Phys. Rev. D",
    volume = "101",
    pages = "074515",
    year = "2020"
}

@article{Djukanovic:2024cmq,
    author = "Djukanovic, Dalibor and von Hippel, Georg and Kuberski, Simon and Meyer, Harvey B. and Miller, Nolan and Ottnad, Konstantin and Parrino, Julian and Risch, Andreas and Wittig, Hartmut",
    title = "{The hadronic vacuum polarization contribution to the muon g {\ensuremath{-}} 2 at long distances}",
    eprint = "2411.07969",
    archivePrefix = "arXiv",
    primaryClass = "hep-lat",
    reportNumber = "CERN-TH-2024-196, MITP-24-080",
    doi = "10.1007/JHEP04(2025)098",
    journal = "JHEP",
    volume = "04",
    pages = "098",
    year = "2025"
}

@article{RBC:2018dos,
    author = {Blum, T. and Boyle, P. A. and G{\"u}lpers, V. and Izubuchi, T. and Jin, L. and Jung, C. and J{\"u}ttner, A. and Lehner, C. and Portelli, A. and Tsang, J. T.},
    collaboration = "RBC, UKQCD",
    title = "{Calculation of the hadronic vacuum polarization contribution to the muon anomalous magnetic moment}",
    eprint = "1801.07224",
    archivePrefix = "arXiv",
    primaryClass = "hep-lat",
    doi = "10.1103/PhysRevLett.121.022003",
    journal = "Phys. Rev. Lett.",
    volume = "121",
    number = "2",
    pages = "022003",
    year = "2018"
}

@article{RBC:2024fic,
    author = "Blum, T. and others",
    collaboration = "RBC, UKQCD",
    title = "{Long-Distance Window of the Hadronic Vacuum Polarization for the Muon g-2}",
    eprint = "2410.20590",
    archivePrefix = "arXiv",
    primaryClass = "hep-lat",
    reportNumber = "CERN-TH-2024-182",
    doi = "10.1103/PhysRevLett.134.201901",
    journal = "Phys. Rev. Lett.",
    volume = "134",
    number = "20",
    pages = "201901",
    year = "2025"
}

@article{Alkofer:2000wg,
    author = "Alkofer, Reinhard and von Smekal, Lorenz",
    title = "{The Infrared behavior of QCD Green's functions: Confinement dynamical symmetry breaking, and hadrons as relativistic bound states}",
    eprint = "hep-ph/0007355",
    archivePrefix = "arXiv",
    reportNumber = "UNITUE-THEP-00-09, FAU-TP3-00-8",
    doi = "10.1016/S0370-1573(01)00010-2",
    journal = "Phys. Rept.",
    volume = "353",
    pages = "281",
    year = "2001"
}

@article{Ball:1980ay,
    author = "Ball, James S. and Chiu, Ting-Wai",
    title = "{Analytic Properties of the Vertex Function in Gauge Theories. 1.}",
    reportNumber = "UU/HEP-80/2",
    doi = "10.1103/PhysRevD.22.2542",
    journal = "Phys. Rev. D",
    volume = "22",
    pages = "2542",
    year = "1980"
}

@article{Miramontes:2022mex,
    author = "Miramontes, {\'A}ngel S. and Alkofer, Reinhard and Fischer, Christian S. and Sanchis-Alepuz, H{\`e}lios",
    title = "{Electromagnetic and strong isospin breaking in light meson masses}",
    eprint = "2202.04618",
    archivePrefix = "arXiv",
    primaryClass = "hep-ph",
    doi = "10.1016/j.physletb.2022.137291",
    journal = "Phys. Lett. B",
    volume = "833",
    pages = "137291",
    year = "2022"
}

@article{Aliberti:2025beg,
 author = "Aliberti, R. and others",
    title = "{The anomalous magnetic moment of the muon in the Standard Model: an update}",
    eprint = "2505.21476",
    archivePrefix = "arXiv",
    primaryClass = "hep-ph",
    reportNumber = "CERN-TH-2025-101, FERMILAB-PUB-25-0344-T, INT-PUB-25-015, IPARCOS-UCM-25-029, KEK Preprint 2025-22, LTH 1403, MITP-25-037, UWThPh 2025-15, UWThPh
  2025-15, ZU-TH 37/25, IPARCOS-UCM-25-029",
    doi = "10.1016/j.physrep.2025.08.002",
    journal = "Phys. Rept.",
    volume = "1143",
    pages = "1--158",
    year = "2025"
}

@unpublished{Eichmann:2025wgs,
    author = "Eichmann, Gernot",
    title = "{Hadron physics with functional methods}",
    eprint = "2503.10397",
    archivePrefix = "arXiv",
    primaryClass = "hep-ph",
    month = "3",
    year = "2025"
}

@unpublised{Huber:2025cbd,
    author = "Huber, Markus Q.",
    title = "{A beginner's guide to functional methods in particle physics}",
    eprint = "2510.18960",
    archivePrefix = "arXiv",
    primaryClass = "hep-ph",
    month = "10",
    year = "2025"
}

@article{Huber:2020ngt,
    author = "Huber, Markus Q. and Fischer, Christian S. and Sanchis-Alepuz, H{\`e}lios",
    title = "{Spectrum of scalar and pseudoscalar glueballs from functional methods}",
    eprint = "2004.00415",
    archivePrefix = "arXiv",
    primaryClass = "hep-ph",
    doi = "10.1140/epjc/s10052-020-08649-6",
    journal = "Eur. Phys. J. C",
    volume = "80",
    number = "11",
    pages = "1077",
    year = "2020"
}

@article{Huber:2021yfy,
    author = "Huber, Markus Q. and Fischer, Christian S. and Sanchis-Alepuz, Helios",
    title = "{Higher spin glueballs from functional methods}",
    eprint = "2110.09180",
    archivePrefix = "arXiv",
    primaryClass = "hep-ph",
    doi = "10.1140/epjc/s10052-021-09864-5",
    journal = "Eur. Phys. J. C",
    volume = "81",
    number = "12",
    pages = "1083",
    year = "2021",
    note = "[Erratum: Eur.Phys.J.C 82, 38 (2022)]"
}

@article{Huber:2025kwy,
    author = "Huber, Markus Q. and Fischer, Christian S. and Sanchis-Alepuz, H{\`e}lios",
    title = "{Apparent convergence in functional glueball calculations}",
    eprint = "2503.03821",
    archivePrefix = "arXiv",
    primaryClass = "hep-ph",
    doi = "10.1140/epjc/s10052-025-14590-3",
    journal = "Eur. Phys. J. C",
    volume = "85",
    number = "8",
    pages = "859",
    year = "2025"
}

@article{Sanchis-Alepuz:2017mir,
    author = "Sanchis-Alepuz, Helios and Alkofer, Reinhard and Fischer, Christian S.",
    title = "{Electromagnetic transition form factors of baryons in the space-like momentum region}",
    eprint = "1707.08463",
    archivePrefix = "arXiv",
    primaryClass = "hep-ph",
    doi = "10.1140/epja/i2018-12465-x",
    journal = "Eur. Phys. J. A",
    volume = "54",
    number = "3",
    pages = "41",
    year = "2018"
}

@article{Sanchis-Alepuz:2015hma,
    author = "Sanchis-Alepuz, Helios and Fischer, Christian S. and Kellermann, Christian and von Smekal, Lorenz",
    title = "{Glueballs from the Bethe-Salpeter equation}",
    eprint = "1503.06051",
    archivePrefix = "arXiv",
    primaryClass = "hep-ph",
    doi = "10.1103/PhysRevD.92.034001",
    journal = "Phys. Rev. D",
    volume = "92",
    pages = "034001",
    year = "2015"
}

@article{Sanchis-Alepuz:2014sca,
    author = "Sanchis-Alepuz, Helios and Fischer, Christian S.",
    title = "{Octet and Decuplet masses: a covariant three-body Faddeev calculation}",
    eprint = "1408.5577",
    archivePrefix = "arXiv",
    primaryClass = "hep-ph",
    doi = "10.1103/PhysRevD.90.096001",
    journal = "Phys. Rev. D",
    volume = "90",
    number = "9",
    pages = "096001",
    year = "2014"
}

@article{Sanchis-Alepuz:2014wea,
    author = "Sanchis-Alepuz, H{\`e}lios and Fischer, Christian S. and Kubrak, Stanislav",
    title = "{Pion cloud effects on baryon masses}",
    eprint = "1401.3183",
    archivePrefix = "arXiv",
    primaryClass = "hep-ph",
    doi = "10.1016/j.physletb.2014.04.031",
    journal = "Phys. Lett. B",
    volume = "733",
    pages = "151--157",
    year = "2014"
}

@article{Miramontes:2025vzb,
    author = "Miramontes, A. S. and Papavassiliou, J. and Pawlowski, J. M.",
    title = "{Electromagnetic properties of heavy-light mesons}",
    eprint = "2508.20631",
    archivePrefix = "arXiv",
    primaryClass = "hep-ph",
    doi = "10.1140/epjc/s10052-025-15121-w",
    journal = "Eur. Phys. J. C",
    volume = "85",
    number = "12",
    pages = "1390",
    year = "2025"
}

@article{Ferreira:2025wpu,
    author = "Ferreira, Mauricio N. and Miramontes, Angel S. and Morgado, Jose M. and Papavassiliou, Joannis and Pawlowski, Jan M.",
    title = "{Pion physics with dressed quark-gluon vertices}",
    eprint = "2512.04853",
    archivePrefix = "arXiv",
    primaryClass = "hep-ph",
    doi = "10.1140/epjc/s10052-026-15487-5",
    journal = "Eur. Phys. J. C",
    volume = "86",
    number = "4",
    pages = "325",
    year = "2026"
}

@unpublished{Hagel:2025ngi,
    author = "Hagel, Stephan and Fischer, Christian S. and Huber, Markus Q. and Yigzaw, Jonathan Y.",
    title = "{Spectra of light and heavy mesons with $J \le 5$ in a relativistic Bethe-Salpeter approach}",
    eprint = "2510.27423",
    archivePrefix = "arXiv",
    primaryClass = "hep-ph",
    month = "10",
    year = "2025"
}

@article{Chen:2023zhh,
    author = "Chen, Chen and Fischer, Christian S. and Roberts, Craig D.",
    title = "{Nucleon-to-{\ensuremath{\Delta}} Axial and Pseudoscalar Transition Form Factors}",
    eprint = "2312.13724",
    archivePrefix = "arXiv",
    primaryClass = "hep-ph",
    reportNumber = "NJU-INP 082/23, USTC-ICTS/PCFT-23-40",
    doi = "10.1103/PhysRevLett.133.131901",
    journal = "Phys. Rev. Lett.",
    volume = "133",
    number = "13",
    pages = "131901",
    year = "2024"
}

@article{Liu:2023reo,
    author = "Liu, Langtian and Fischer, Christian S.",
    title = "{Space-like electromagnetic form factors of lambda- and sigma-baryons from quark-diquark Faddeev equations}",
    eprint = "2311.13269",
    archivePrefix = "arXiv",
    primaryClass = "hep-ph",
    doi = "10.1140/epja/s10050-024-01283-w",
    journal = "Eur. Phys. J. A",
    volume = "60",
    number = "4",
    pages = "84",
    year = "2024"
}

@article{Eichmann:2016hgl,
    author = "Eichmann, Gernot and Fischer, Christian S. and Sanchis-Alepuz, Helios",
    title = "{Light baryons and their excitations}",
    eprint = "1607.05748",
    archivePrefix = "arXiv",
    primaryClass = "hep-ph",
    doi = "10.1103/PhysRevD.94.094033",
    journal = "Phys. Rev. D",
    volume = "94",
    number = "9",
    pages = "094033",
    year = "2016"
}

@article{Sanchis-Alepuz:2015qra,
    author = "Sanchis-Alepuz, Helios and Williams, Richard",
    title = "{Probing the quark{\textendash}gluon interaction with hadrons}",
    eprint = "1504.07776",
    archivePrefix = "arXiv",
    primaryClass = "hep-ph",
    doi = "10.1016/j.physletb.2015.08.067",
    journal = "Phys. Lett. B",
    volume = "749",
    pages = "592--596",
    year = "2015"
}

@article{Hoffer:2024fgm,
    author = "Hoffer, Joshua and Eichmann, Gernot and Fischer, Christian S.",
    title = "{Structure of open-flavor four-quark states in the charm and bottom region}",
    eprint = "2409.05779",
    archivePrefix = "arXiv",
    primaryClass = "hep-ph",
    doi = "10.1103/PhysRevD.111.054028",
    journal = "Phys. Rev. D",
    volume = "111",
    number = "5",
    pages = "054028",
    year = "2025"
}

@article{Hoffer:2024alv,
    author = "Hoffer, Joshua and Eichmann, Gernot and Fischer, Christian S.",
    title = "{Hidden-flavor four-quark states in the charm and bottom region}",
    eprint = "2402.12830",
    archivePrefix = "arXiv",
    primaryClass = "hep-ph",
    doi = "10.1103/PhysRevD.109.074025",
    journal = "Phys. Rev. D",
    volume = "109",
    number = "7",
    pages = "074025",
    year = "2024"
}

@article{Santowsky:2020pwd,
    author = "Santowsky, Nico and Eichmann, Gernot and Fischer, Christian S. and Wallbott, Paul C. and Williams, Richard",
    title = "{$\sigma$-meson: Four-quark versus two-quark components and decay width in a Bethe-Salpeter approach}",
    eprint = "2007.06495",
    archivePrefix = "arXiv",
    primaryClass = "hep-ph",
    doi = "10.1103/PhysRevD.102.056014",
    journal = "Phys. Rev. D",
    volume = "102",
    number = "5",
    pages = "056014",
    year = "2020"
}

@article{Wallbott:2019dng,
    author = "Wallbott, Paul C. and Eichmann, Gernot and Fischer, Christian S.",
    title = "{$X(3872)$ as a four-quark state in a Dyson-Schwinger/Bethe-Salpeter approach}",
    eprint = "1905.02615",
    archivePrefix = "arXiv",
    primaryClass = "hep-ph",
    doi = "10.1103/PhysRevD.100.014033",
    journal = "Phys. Rev. D",
    volume = "100",
    number = "1",
    pages = "014033",
    year = "2019"
}

@article{Heupel:2012ua,
    author = "Heupel, Walter and Eichmann, Gernot and Fischer, Christian S.",
    title = "{Tetraquark Bound States in a Bethe-Salpeter Approach}",
    eprint = "1206.5129",
    archivePrefix = "arXiv",
    primaryClass = "hep-ph",
    doi = "10.1016/j.physletb.2012.11.009",
    journal = "Phys. Lett. B",
    volume = "718",
    pages = "545--549",
    year = "2012"
}

@article{Eichmann:2024glq,
    author = "Eichmann, Gernot and Fischer, Christian S. and Haeuser, Tim and Regenfelder, Oliver",
    title = "{Axial-vector and scalar contributions to hadronic light-by-light scattering}",
    eprint = "2411.05652",
    archivePrefix = "arXiv",
    primaryClass = "hep-ph",
    doi = "10.1140/epjc/s10052-025-14055-7",
    journal = "Eur. Phys. J. C",
    volume = "85",
    number = "4",
    pages = "445",
    year = "2025"
}

@article{Xing:2025iab,
    author = "Xing, Zanbin and Chang, Lei and Raya, Kh{\'e}pani",
    title = "{Electromagnetic structure of axial-vector mesons and implications for the muon g-2}",
    eprint = "2509.17311",
    archivePrefix = "arXiv",
    primaryClass = "hep-ph",
    doi = "10.1103/xbls-4zqr",
    journal = "Phys. Rev. D",
    volume = "113",
    number = "3",
    pages = "034010",
    year = "2026"
}

@article{Raya:2019dnh,
    author = "Raya, Kh{\'e}pani and Bashir, Adnan and Roig, Pablo",
    title = "{Contribution of neutral pseudoscalar mesons to $a_\mu^{HLbL}$ within a Schwinger-Dyson equations approach to QCD}",
    eprint = "1910.05960",
    archivePrefix = "arXiv",
    primaryClass = "hep-ph",
    doi = "10.1103/PhysRevD.101.074021",
    journal = "Phys. Rev. D",
    volume = "101",
    number = "7",
    pages = "074021",
    year = "2020"
}

@article{Goecke:2011pe,
    author = "Goecke, Tobias and Fischer, Christian S. and Williams, Richard",
    title = "{Leading-order calculation of hadronic contributions to the muon $g-2$ using the Dyson-Schwinger approach}",
    eprint = "1107.2588",
    archivePrefix = "arXiv",
    primaryClass = "hep-ph",
    doi = "10.1016/j.physletb.2011.09.019",
    journal = "Phys. Lett. B",
    volume = "704",
    pages = "211--217",
    year = "2011"
}

@article{Fischer:2008wy,
    author = "Fischer, Christian S. and Williams, Richard",
    title = "{Beyond the rainbow: Effects from pion back-coupling}",
    eprint = "0808.3372",
    archivePrefix = "arXiv",
    primaryClass = "hep-ph",
    doi = "10.1103/PhysRevD.78.074006",
    journal = "Phys. Rev. D",
    volume = "78",
    pages = "074006",
    year = "2008"
}

@article{Fischer:2008sp,
    author = "Fischer, Christian S. and Nickel, Dominik and Williams, Richard",
    title = "{On Gribov's supercriticality picture of quark confinement}",
    eprint = "0807.3486",
    archivePrefix = "arXiv",
    primaryClass = "hep-ph",
    reportNumber = "MIT-CTP-3964",
    doi = "10.1140/epjc/s10052-008-0821-1",
    journal = "Eur. Phys. J. C",
    volume = "60",
    pages = "47--61",
    year = "2009"
}

@article{Greynat:2022geu,
    author = "Greynat, David and de Rafael, Eduardo",
    title = "{Hadronic vacuum polarization and the MUonE proposal}",
    eprint = "2202.10810",
    archivePrefix = "arXiv",
    primaryClass = "hep-ph",
    doi = "10.1007/JHEP05(2022)084",
    journal = "JHEP",
    volume = "05",
    pages = "084",
    year = "2022"
}

@article{Xu:2022kng,
    author = "Xu, Zhen-Ni and Yao, Zhao-Qian and Qin, Si-Xue and Cui, Zhu-Fang and Roberts, Craig D.",
    title = "{Bethe{\textendash}Salpeter kernel and properties of strange-quark mesons}",
    eprint = "2208.13903",
    archivePrefix = "arXiv",
    primaryClass = "hep-ph",
    reportNumber = "NJU-INP 065/22",
    doi = "10.1140/epja/s10050-023-00951-7",
    journal = "Eur. Phys. J. A",
    volume = "59",
    number = "3",
    pages = "39",
    year = "2023"
}

@unpublished{Ferreira:2026gbe,
    author = "Ferreira, M. N. and Miramontes, A. S. and Morgado, J. M. and Papavassiliou, J.",
    title = "{Light mesons in the symmetric-vertex approximation}",
    eprint = "2604.07221",
    archivePrefix = "arXiv",
    primaryClass = "hep-ph",
    month = "4",
    year = "2026"
}

@article{Miramontes:2024fgo,
    author = "Miramontes, Angel S. and Raya, K. and Bashir, A. and Roig, P. and Paredes-Torres, G.",
    title = "{Radially excited pion: electromagnetic form factor and the box contribution to the muon's $g-2$}",
    eprint = "2411.02218",
    archivePrefix = "arXiv",
    primaryClass = "hep-ph",
    doi = "10.1088/1674-1137/add259",
    journal = "Chin. Phys. C",
    volume = "49",
    number = "8",
    pages = "083108",
    year = "2025"
}

@article{Goecke:2010if,
    author = "Goecke, Tobias and Fischer, Christian S. and Williams, Richard",
    title = "{Hadronic light-by-light scattering in the muon g-2: a Dyson-Schwinger equation approach}",
    eprint = "1012.3886",
    archivePrefix = "arXiv",
    primaryClass = "hep-ph",
    doi = "10.1103/PhysRevD.83.094006",
    journal = "Phys. Rev. D",
    volume = "83",
    pages = "094006",
    year = "2011",
    note = "[Erratum: Phys.Rev.D 86, 099901 (2012)]"
}

@article{Aoyama:2012wk,                                    
  author        = "Aoyama, Tatsumi and Hayakawa, Masashi and Kinoshita,                                                                 Toichiro and Nio, Makiko",               
  title         = "{Complete Tenth-Order QED Contribution to the Muon $g-2$}",                                        
  journal       = "Phys. Rev. Lett.",                      
  volume        = "109",
  year          = "2012",
  pages         = "111808",
  doi           = "10.1103/PhysRevLett.109.111808",        
  eprint        = "1205.5370",
  archiveprefix = "arXiv",
  primaryclass  = "hep-ph",
  reportnumber  = "RIKEN-QHP-26",
  slaccitation  = "%%CITATION = ARXIV:1205.5370;%%"        
}

@article{Volkov:2019phy,
  author        = "Volkov, Sergey",
  title         = "{Calculating the five-loop QED contribution to the
                  electron anomalous magnetic moment: Graphs without lepton
                  loops}",
  journal       = "Phys. Rev. D",
  volume        = "100",
  year          = "2019",
  number        = "9",
  pages         = "096004",
  doi           = "10.1103/PhysRevD.100.096004",
  eprint        = "1909.08015",
  archiveprefix = "arXiv",
  primaryclass  = "hep-ph",
  slaccitation  = "%%CITATION = ARXIV:1909.08015;%%"       
}

@article{Volkov:2024yzc,
    author = "Volkov, Sergey",
    title = "{Calculation of the total 10th order QED contribution to the electron magnetic moment}",
    eprint = "2404.00649",
    archivePrefix = "arXiv",
    primaryClass = "hep-ph",
    doi = "10.1103/PhysRevD.110.036001",
    journal = "Phys. Rev. D",
    volume = "110",
    number = "3",
    pages = "036001",
    year = "2024"
}

@article{Aoyama:2024aly,
    author = "Aoyama, Tatsumi and Hayakawa, Masashi and Hirayama, Akira and Nio, Makiko",
    title = "{Verification of the tenth-order QED contribution to the anomalous magnetic moment of the electron from diagrams without fermion loops}",
    eprint = "2412.06473",
    archivePrefix = "arXiv",
    primaryClass = "hep-ph",
    doi = "10.1103/PhysRevD.111.L031902",
    journal = "Phys. Rev. D",
    volume = "111",
    number = "3",
    pages = "L031902",
    year = "2025"
}

@article{Parker:2018vye,
  author        = "Parker, Richard H. and Yu, Chenghui and Zhong, Weicheng
                  and Estey, Brian and M{\"u}ller, Holger",
  title         = "{Measurement of the fine-structure constant as a test of
                  the Standard Model}",
  journal       = "Science",
  volume        = "360",
  year          = "2018",
  pages         = "191",
  doi           = "10.1126/science.aap7706",
  eprint        = "1812.04130",
  archiveprefix = "arXiv",
  primaryclass  = "physics.atom-ph",
  slaccitation  = "%%CITATION = ARXIV:1812.04130;%%"       
}

@article{Morel:2020dww,
    author = {Morel, L\'eo and Yao, Zhibin and Clad\'e, Pierre and Guellati-Kh\'elifa, Sa\"\i{}da},
    title = "{Determination of the fine-structure constant with an accuracy of 81 parts per trillion}",
    doi = "10.1038/s41586-020-2964-7",
    journal = "Nature",
    volume = "588",
    number = "7836",
    pages = "61--65",
    year = "2020"
}

@article{Fan:2022eto,
    author = "Fan, X. and Myers, T. G. and Sukra, B. A. D. and Gabrielse, G.",
    title = "{Measurement of the Electron Magnetic Moment}",
    eprint = "2209.13084",
    archivePrefix = "arXiv",
    primaryClass = "physics.atom-ph",
    doi = "10.1103/PhysRevLett.130.071801",
    journal = "Phys. Rev. Lett.",
    volume = "130",
    number = "7",
    pages = "071801",
    year = "2023"
}

@article{Czarnecki:2002nt,
  author        = "Czarnecki, Andrzej and Marciano, William J. and
                  Vainshtein, Arkady",
  title         = "{Refinements in electroweak contributions to the muon
                  anomalous magnetic moment}",
  journal       = "Phys. Rev. D",
  volume        = "67",
  year          = "2003",
  pages         = "073006",
  doi           = "10.1103/PhysRevD.67.073006",
  note          = "[Erratum: Phys. Rev. D {\bf 73}, 119901 (2006)]",
  eprint        = "hep-ph/0212229",
  archiveprefix = "arXiv",
  primaryclass  = "hep-ph",
  reportnumber  = "ALBERTA-THY-18-02, BNL-HET-02-25, TPI-MINN-02-46,
                  UMN-TH-2119-02",
  slaccitation  = "%%CITATION = HEP-PH/0212229;%%"
}

@article{Gnendiger:2013pva,
  author        = "Gnendiger, C. and St{\"o}ckinger, D. and
                  St{\"o}ckinger-Kim, H.",
  title         = "{The electroweak contributions to $(g-2)_\mu$ after the
                  Higgs boson mass measurement}",
  journal       = "Phys. Rev. D",
  volume        = "88",
  year          = "2013",
  pages         = "053005",
  doi           = "10.1103/PhysRevD.88.053005",
  eprint        = "1306.5546",
  archiveprefix = "arXiv",
  primaryclass  = "hep-ph",
  slaccitation  = "%%CITATION = ARXIV:1306.5546;%%"        
}

@article{Ludtke:2024ase,
    author = {L\"udtke, Jan and Procura, Massimiliano and Stoffer, Peter},
    title = "{Dispersion relations for the hadronic VVA correlator}",
    eprint = "2410.11946",
    archivePrefix = "arXiv",
    primaryClass = "hep-ph",
    reportNumber = "PSI-PR-24-18, UWThPh 2024-20, ZU-TH 49/24",
    doi = "10.1007/JHEP04(2025)130",
    journal = "JHEP",
    volume = "04",
    pages = "130",
    year = "2025"
}

@article{Hoferichter:2025yih,
    author = {Hoferichter, Martin and L\"udtke, Jan and Naterop, Luca and Procura, Massimiliano and Stoffer, Peter},  
    title = "{Improved evaluation of the electroweak contribution to muon $g-2$}",
    eprint = "2503.04883",
    archivePrefix = "arXiv",
    primaryClass = "hep-ph",
    reportNumber = "PSI-PR-25-04, UWThPh 2025-8, ZU-TH 10/25",
    doi = "10.1103/PhysRevLett.134.201801",
    journal = "Phys. Rev. Lett.",
    volume = "134",
    number = "20",
    pages = "201801",
    year = "2025"
}

@article{Giusti:2019xct,
  author        = "Giusti, D. and Lubicz, V. and Martinelli, G. and
                  Sanfilippo, F. and Simula, S.",
  collaboration = {ETM},
  title         = "{Electromagnetic and strong isospin-breaking corrections
                  to the muon $g - 2$ from Lattice QCD+QED}",
  journal       = "Phys. Rev. D",
  volume        = "99",
  year          = "2019",
  number        = "11",
  pages         = "114502",
  doi           = "10.1103/PhysRevD.99.114502",
  eprint        = "1901.10462",
  archiveprefix = "arXiv",
  primaryclass  = "hep-lat",
  slaccitation  = "%%CITATION = ARXIV:1901.10462;%%"       
}

@article{Wang:2022lkq,
    author = "Wang, Gen and Draper, Terrence and Liu, Keh-Fei and Yang, Yi-Bo",
    collaboration = "$\chi$QCD",
    title = "{Muon g-2 with overlap valence fermions}",    
    eprint = "2204.01280",
    archivePrefix = "arXiv",
    primaryClass = "hep-lat",
    doi = "10.1103/PhysRevD.107.034513",
    journal = "Phys. Rev. D",
    volume = "107",
    number = "3",
    pages = "034513",
    year = "2023"
}

@article{Aubin:2022hgm,
    author = "Aubin, Christopher and Blum, Thomas and Golterman, Maarten and Peris, Santiago",
    title = "{Muon anomalous magnetic moment with staggered fermions: Is the lattice spacing small enough?}",
    eprint = "2204.12256",
    archivePrefix = "arXiv",
    primaryClass = "hep-lat",
    doi = "10.1103/PhysRevD.106.054503",
    journal = "Phys. Rev. D",
    volume = "106",
    number = "5",
    pages = "054503",
    year = "2022"
}

@article{Ce:2022kxy,
    author = "C\`e, Marco and others",
    title = "{Window observable for the hadronic vacuum polarization contribution to the muon $g-2$ from lattice QCD}",
    eprint = "2206.06582",
    archivePrefix = "arXiv",
    primaryClass = "hep-lat",
    reportNumber = "MITP-22-038, CERN-TH-2022-098, DESY-22-105",
    doi = "10.1103/PhysRevD.106.114502",
    journal = "Phys. Rev. D",
    volume = "106",
    number = "11",
    pages = "114502",
    year = "2022"
}

@article{ExtendedTwistedMass:2022jpw,
    author = "Alexandrou, C. and others",
    collaboration = "ETM",
    title = "{Lattice calculation of the short and intermediate time-distance hadronic vacuum polarization contributions to the muon magnetic moment using twisted-mass fermions}",
    eprint = "2206.15084",
    archivePrefix = "arXiv",
    primaryClass = "hep-lat",
    doi = "10.1103/PhysRevD.107.074506",
    journal = "Phys. Rev. D",
    volume = "107",
    number = "7",
    pages = "074506",
    year = "2023"
}

@article{RBC:2023pvn,
    author = "Blum, T. and others",
    collaboration = "RBC, UKQCD",
    title = "{Update of Euclidean windows of the hadronic vacuum polarization}",
    eprint = "2301.08696",
    archivePrefix = "arXiv",
    primaryClass = "hep-lat",
    doi = "10.1103/PhysRevD.108.054507",
    journal = "Phys. Rev. D",
    volume = "108",
    number = "5",
    pages = "054507",
    year = "2023"
}

@article{Kuberski:2024bcj,
    author = "Kuberski, Simon and C\`e, Marco and von Hippel, Georg and Meyer, Harvey B. and Ottnad, Konstantin and Risch, Andreas and Wittig, Hartmut",
    title = "{Hadronic vacuum polarization in the muon g \ensuremath{-} 2: the short-distance contribution from lattice QCD}",
    eprint = "2401.11895",
    archivePrefix = "arXiv",
    primaryClass = "hep-lat",
    reportNumber = "MITP-24-011, CERN-TH-2024-011",        
    doi = "10.1007/JHEP03(2024)172",
    journal = "JHEP",
    volume = "03",
    pages = "172",
    year = "2024"
}

@article{Boccaletti:2024guq,
    author = "Boccaletti, A. and others",
    title = "{Hybrid calculation of hadronic vacuum polarization in muon g {\ensuremath{-}} 2 to 0.48{\%}}",
    eprint = "2407.10913",
    archivePrefix = "arXiv",
    primaryClass = "hep-lat",
    doi = "10.1038/s41586-026-10449-z",
    journal = "Nature",
    volume = "653",
    number = "8114",
    pages = "373--377",
    year = "2026"
}

@article{Spiegel:2024dec,
    author = "Spiegel, Sebastian and Lehner, Christoph",
    title = "{High-precision continuum limit study of the HVP short-distance window}",
    eprint = "2410.17053",
    archivePrefix = "arXiv",
    primaryClass = "hep-lat",
    doi = "10.1103/mj3d-yq87",
    journal = "Phys. Rev. D",
    volume = "111",
    number = "11",
    pages = "114517",
    year = "2025"
}

@article{MILC:2024ryz,
    author = "Bazavov, Alexei and others",
    collaboration = "Fermilab Lattice, HPQCD, MILC",       
    title = "{Hadronic vacuum polarization for the muon g-2 from lattice QCD: Complete short and intermediate windows}",
    eprint = "2411.09656",
    archivePrefix = "arXiv",
    primaryClass = "hep-lat",
    reportNumber = "FERMILAB-PUB-24-0835-T",
    doi = "10.1103/PhysRevD.111.094508",
    journal = "Phys. Rev. D",
    volume = "111",
    number = "9",
    pages = "094508",
    year = "2025"
}

@article{FermilabLatticeHPQCD:2024ppc,
    author = "Bazavov, Alexei and others",
    collaboration = "Fermilab Lattice, HPQCD, MILC",
    title = "{Hadronic Vacuum Polarization for the Muon $g-2$ from Lattice QCD: Long-Distance and Full Light-Quark Connected Contribution}",
    eprint = "2412.18491",
    archivePrefix = "arXiv",
    primaryClass = "hep-lat",
    reportNumber = "FERMILAB-PUB-24-0957-T",
    doi = "10.1103/d583-yhfs",
    journal = "Phys. Rev. Lett.",
    volume = "135",
    number = "1",
    pages = "011901",
    year = "2025"
}

@article{Keshavarzi:2019abf,
  author        = "Keshavarzi, Alexander and Nomura, Daisuke and Teubner,
                  Thomas",
  title         = "{The $g-2$ of charged leptons, $\alpha(M_Z^2)$ and the
                  hyperfine splitting of muonium}",        
  journal       = "Phys. Rev. D",
  volume        = "101",
  year          = "2020",
  pages         = "014029",
  doi           = "10.1103/PhysRevD.101.014029",
  eprint        = "1911.00367",
  archiveprefix = "arXiv",
  primaryclass  = "hep-ph",
  slaccitation  = "%%CITATION = ARXIV:1911.00367;%%"       
}

@article{ParticleDataGroup:2026aaa,
    author = "Takahashi, F. and others",
    collaboration = "Particle Data Group",
    title = "{Review of Particle Physics}",
    doi = "10.1142/S0217751X26300115",
    journal = "Int. J. Mod. Phys. A",
    volume = "41",
    pages = "2630011",
    year = "2026"
}

\end{document}